\documentclass[table, dvipsnames,twocolumn]{aastex631}
\usepackage{amsmath}
\usepackage{graphicx}
\graphicspath{{images/}}
\usepackage[utf8]{inputenc}
\usepackage{csquotes}
\usepackage{float}

\usepackage{enumerate}
\usepackage{verbatim}
\usepackage{todonotes}
\usepackage{hyperref}
\usepackage[capitalize, noabbrev]{cleveref}

\pdfoutput=1

\crefname{section}{\S}{\S\S}

\makeatletter
\usepackage{etoolbox}
\patchcmd\H@refstepcounter{\protected@edef}{\protected@xdef}{}{}
\makeatother

\newcommand{\myemail}{epadill7@jhu.edu}

\newcommand{\STScI}{Space Telescope Science Institute, Baltimore, MD 21218, USA}
\newcommand{\JHU}{Department of Physics and Astronomy, Johns Hopkins University, Baltimore, MD 21218, USA}

\received{2025}
\accepted{}

\begin{document}

\title{Revisiting the Mass Step: Environmental Dependence of Type Ia Supernovae in Low-Metallicity Host Galaxies}

\shorttitle{}
\shortauthors{Padilla~Gonzalez et al.}

\author[0000-0003-0209-9246]{Padilla~Gonzalez, E.}
\affiliation{\JHU}
\email{\myemail}

\author[0000-0002-7593-8584]{Bhavin, A.~J.}
\affiliation{\JHU}

\author[0000-0002-7756-4440]{Strolger, L.~G.}
\affiliation{\STScI}

\author{Khatri, B.}
\affiliation{Department of Astronomy, Case Western Reserve University,10900 Euclid Avenue, Cleveland, OH 44106 }

\author{Rest, F.}
\affiliation{\STScI}
\affiliation{\JHU}

\author[0000-0002-4410-5387]{Rest, A.}
\affiliation{\STScI}

\author[0000-0002-1873-8973]{Rose, B.}
\affiliation{Department of Physics and Astronomy, Baylor University, One Bear Place 97316, Waco, TX 76798-7316}

\author[0000-0001-5281-731X]{Angulo, R.}
\affiliation{\JHU}

\author[0000-0003-4263-2228]{Coulter, D.}
\affiliation{\JHU}

\author[0000-0002-7566-6080]{Derkacy, J.~ M.}
\affiliation{\STScI}

\author[0000-0003-2238-1572]{Fox, O.}
\affiliation{\STScI}

\author[0000-0002-2361-7201]{Pierel, J.~D.~R.}
\affiliation{\STScI}

\author[0000-0002-2798-2943]{Shukawa,K.}
\affiliation{\JHU}

\author[0000-0002-9301-5302]{Shahbandeh, M.}
\affiliation{\STScI}

\author[0000-0003-2445-3891]{Siebert, M.}
\affiliation{\STScI}

\author[0000-0003-2037-4619]{Larison, C.}
\affiliation{\STScI}

\author[0000-0002-5060-1379]{Griggio, M.}
\affiliation{\STScI}

\begin{abstract}
Despite the tremendous impact of Type Ia supernovae (SNe~Ia) on the field of cosmology, their underlying physics are still poorly understood. Studies have found an intriguing correlation between standardized Type Ia supernova (SN~Ia) luminosities and host galaxy masses, commonly referred to as the \textquote{mass step}. SNe~Ia in massive galaxies appear systematically brighter than in lower-mass, star-forming hosts after standardization. However, previous analyses utilize host galaxy mass estimates derived from largely optical data alone, and assume parametric forms for host star formation histories (SFHs), both of which are known to misestimate galaxy stellar masses. In this work we re-examine the mass-step relation with a sample of SN~Ia host galaxies complete in broadband optical ($\sim$3000\AA{} - 1$\mu m$) and near-IR ($\sim1-1.8\mu m$), and in some cases with data up to the mid-IR (to $\sim4.5\mu m$), using {\tt prospector} to derive nonparametric SFHs. We find that while the masses for these sample galaxies have indeed been underestimated, the overall trend in SN~Ia luminosity versus host mass remains largely unchanged. However, we also uncover an environmental metallicity-dependent trend, in which low metallicity galaxies (Z $<$ -1 $Z_\odot$) may drive much of the observed SN~Ia luminosity–mass correlation, suggesting that chemical composition of the SN environment plays a central role in shaping the standardized brightness of SNe~Ia.
\end{abstract}


\section{Introduction}
\label{sec:intro}
Type Ia supernovae (SNe~Ia) are powerful probes for measuring the expansion of the Universe. Their use led to the discovery of the accelerated expansion of the universe \citep{Riess1998, Perlmutter1999}. This class of objects are excellent tools for measuring distances as they are standardizable due to the relationship between their peak brightness, light curve width \citep{Phillips1993} and color relationship \citep{Riess1996, Tripp1998}, which together account for $\sim $0.14 mag of brightness variation. Although some of the remaining scatter may be attributed to observational uncertainties, such as the \textquote{mass-step} (described further in Section~\ref{sec:hr}) and other bias corrections, another is \textquote{residual dispersion} of $\sim $0.08-0.10 mag \cite{Brout2019}. This suggests that SNe~Ia may be standardizable up to a limit, or there may be external factors which are not fully represented in the literature. This can include explosion mechanisms, progenitor systems —whether single- or double-degenerate \citep{Maoz2014, Livio&mazzali2018} —and/or environmental effects \citep{ maguire2017, Rigault2013, Jones2018}.
 A comprehensive understanding of the SNe~Ia intrinsic scatter and its underlying characterization remains elusive. 

Global and local properties of SNe~Ia host galaxies such as stellar mass, star formation rate (SFR), and stellar population age have all shown to correlate with the distance modulus residuals after standardization, but its physical origin remains ambiguous \citep{Sullivan2010, Kelly2010, Lampeitl2010, childress2013, Rose2019}. However, interpreting these trends relies critically on accurate characterization of host-galaxy properties, which is currently limited by the paucity of rest-frame near-IR photometry and the constraints of widely used spectral energy distribution (SED)-fitting tools.


A physical implication of the \textquote{mass-step}-- the trend observed between standardized residuals and host-galaxy stellar mass-- is that the explosion properties of SNe~Ia depend on the progenitor host environment. This in turn, suggests that our current standardization does not fully capture the underlying physics governing the luminosity. Another implication is that SNe~Ia have at least two principal environments, one from older and possibly metal-rich galaxies (traced by high-mass galaxies) or younger, metal-poor galaxies~\citep[traced by low-mass galaxies, ][]{Wojtak2023, Larison2024}. 

Yet, it is also known that galaxy characteristics, i.e., their stellar masses, star-formation histories (SFHs), and star-formation rates (SFRs), are difficult to estimate from broadband optical photometry alone, as the contributions of low-mass stars to the total mass budget, and the full effects of attenuation, are not well accounted for at these wavelengths. These are better determined by including broadband photometry in the near-IR (and mid-IR)~\citep{Pforr2012, Pforr2013, Mobasher2015}. Nonetheless, limited work has been done to evaluate the properties of the SNe~Ia hosts beyond the optical, with few exceptions~\citep[e.g.,][]{Uddin2020, Peterson2024}. Furthermore, previous SN host galaxy SED fits assume simple parametric SFH. These models typically adopt exponentially declining SFH forms, where SFH $\propto$ $t \exp{(-t/\tau)}$ and often underestimate stellar masses for dusty and starbursting galaxies \citep{Mitchell2013, Leja2019}. These commonly used parametric SFH models allow for a linear rise in star formation at early times and an exponential decay at later times, forcing the late-time SFRs to be dependent on early time SFR, biasing stellar ages and SFRs \citep{Lower2020}, which is especially problematic for early-type galaxy with recent star formation.  Furthermore, these parametric models are known to disagree with the directly observed cosmic star formation density rate \citep{Madau_dickinson2014ARA&A..52..415M, behroozi2019, Leja2020}. 

Finally, exponential parametric SFHs have been shown to systematically under predict the SFRs and thus total stellar masses of galaxies at $z>2$~\citep{Santini2009, wuyts2011}.  For more precise mass estimates, using a non-parametric SFHs, which employs a piecewise constant SFH model within specified age bins, provides better assumptions and avoids underestimating stellar masses~\citep{Leja2019, Carnall2019, Lower2020}. By studying SNe~Ia and their physical implications the hope is to reduce potential biases to enable more precise measurements of the Hubble constant $H_{0}$. 


In this paper, we recover archival optical ($\sim$3000\AA{} - 1$\mu m$), near-IR ($\sim1-1.8\mu m$), and in some cases, to mid-IR ($< 4.0\mu m$) host-photometry for a well-calibrated subsample of SN~Ia and their host galaxies~\cite[Pantheon+ hereafter]{Brout2022}. We compare newly derived host galaxy SED fits directly with previous Pantheon+ results. We compile host-galaxy photometry from the COSMOS2020 catalog \citep{cosmos2020}, the S82-MGC catalog in the Stripe 82 region \citep{Bundy2015}, and the Carnegie Supernovae Project \citep[CSP;][]{Uddin2020}. Furthermore, we introduce an improved non-parametric SED-fitting framework that leverages the photometry for the hosts galaxies to more robustly estimate key SNe~Ia host properties. With these new mass estimates, we re-examine Pantheon+ mass-step relation to assess whether the residuals truly depend on host-galaxy stellar mass or other physical host properties. 


In Section~\ref{sec:hr} we define the Hubble residuals and mass step fitting in our analysis. In Section~\ref{sec:data} we discuss the sources of archival data for this new analysis, and corrections applied to derive accurate SEDs fits. In Section~\ref{sec:stellarmass}, we compare the derived stellar masses with the Pantheon+ results. In Section~\ref{sec:trends_hr} we discuss the mass-step and observed physical and non-physical trends with the Hubble residuals. In Section~\ref{sec:cosmology}, we discuss the impact of removing HR trends on cosmology by comparing them to the full-sample. Finally, in Section~\ref{sec:conclusion}, we discuss our results and conclusion.

\section{Hubble Residuals and Mass Step}\label{sec:hr}
The Hubble residual (HR) is defined as  $\mu_{obs}$ -  $\mu_{model}$, where $\mu_{model}$ is determined assuming a standard $\Lambda$CDM cosmology, with $\Omega_{m} = 0.27$ and $H_{0} =74$ (km/s/Mpc).  While $\mu_{obs}$ is estimated from the fitted SNe~Ia lightcurve parameters with \texttt{SALT2}~\citep{SALT22007} by,  
\begin{equation}
    \mu_{obs}= m_{B} + \alpha x_{1} - \beta c - M_{B}.
    \label{mu_obs}
\end{equation}
\noindent The best-fit parameters were taken directly from the Pantheon+ literature \citep{Scolnic2021},  where apparent peak magnitude, nominally in the $B-$band ($m_B$), light-curve stretch ($x1$) and color ($c$) are fitted and $\alpha$, $\beta$, and absolute magnitude ($M_{B}$) are derived constants.


As described in \cite{Brout2022b}, the Pantheon+ dataset consists of results from 18 different survey samples, re-calibrated to a consistent photometric system. Redshifts were corrected against potential local void and measurement biases, and peculiar motions. The standardization of the SN~Ia light curves and fits were performed in \cite{Scolnic2022}, using \citep[SNANA;][]{Kessler2009} to fit the SALT2 \citep{SALT22007} model to the data. 

\cite{Scolnic2022} derive the distance modulus from the light-curve parameters, the following relation from~\citep{Tripp1998},

\begin{equation}
    \mu'_{obs} = m_{B} + \alpha x_{1} - \beta c - 
    \rm M_B - \delta_{\mu - bias},
    \label{eqn:muobs}
\end{equation}
where the $\delta_{\mu - bias}$ accounts for selection effects and the \textquote{mass-step} correction, $\delta_{\mu - host}$. However, since in this analysis we are interested in revisiting the SNe~Ia brightness and host galaxy mass relation, we remove the $\delta_{\mu - host}$ from Equation~\ref{eqn:muobs} assuming,

\begin{equation}
    \delta_{\mu - host} = \gamma\{1+e^{-[\log(M/M_{\odot})-M_{\rm step}]/\tau}\}^{-1}  - \gamma/2,
    \label{delta_mu}
\end{equation}
as defined in~\cite{Brout2021} and \cite{Popovic2021}. Here, $M$ is the mass of the galaxy, $M_{\rm step}$  is the point of transition in the mass-step, in log space, and is set to 10. The extent of the transition ($\gamma$) and width of the transition ($\tau$) are set to $0.0166$ and $0.01$, respectively. In this paper, we adopt the observed distance modulus, $\mu_{obs}$, which excludes  $\delta_{\mu - bias}$ and the effects of the mass-step, as
\begin{equation}
    \centering
    \mu_{obs}= \mu'_{obs} +  \delta_{\mu - bias},
\end{equation}
while keeping all other Pantheon+ SN~Ia fitted parameters unchanged.

\section{Data}\label{sec:data}
Our subset of events and host galaxies was taken from the Pantheon+ catalog and cross-matched with catalogs with extensive optical and near-IR data. We obtained most of the host galaxy photometry from various source catalogs, including COSMOS2020 \citep{cosmos2020}, S82-MGC \citep{Bundy2015}, CSP-I \citep{Uddin2020, Krisciunas2017}, and SDSS-DR16 \citep{sdssdr16}. A small subset of galaxy photometry was determined from \emph{HST} archival imaging, retrieved from the Mikulski Archive for Space Telescopes \citep[MAST;][]{Marston2018}. 



\subsection{Multi-wavelength photometry for host galaxies}
Consistently measuring photometry using data from multiple surveys is challenging, especially for extended sources over a range of resolutions and detection limits. Therefore, we largely adopted photometry from carefully curated catalogs COSMOS2020, S82-MGC, and CSP-I, where various tests and validation efforts were undertaken, and value-added products were derived.  For the COSMOS2020 catalog from \citet{cosmos2020}, astrometric solutions were recomputed using the Gaia catalog, images were convolved for consistent Moffat-shaped PSF profiles for photometry across various bands. The photometry was validated through color checks, literature comparisons, and different extraction methods, including both aperture photometric methods and the new profile-fitting photometric extraction tool, \texttt{The Farmer} \citep{cosmos2020, Weaver2023}. For the S82-MGC catalog, photometry was validated by matching filters across a consistent PSF, applying cuts in limiting depths, improving star galaxy separation by using  ($J - K$) vs. ($g - i$) colors to separate galaxies from the stellar locus, and geometric masking \citep{Bundy2015}. The CSP-I catalog ensured consistent photometry by comparing their optical and NIR photometry to (SDSS) model magnitudes available from the SDSS Sky Server \citep{sdssdr16} and the Two Micron All Sky Survey \citep[2MASS;][]{2mass2006} source catalog, respectively. Finally, the remainder of data were obtained from the MAST and SDSS-DR16 catalogs, both the images and apertures were inspected individually to ensure consistent galaxy identification and flux across the optical and NIR bands. The magnitudes used from the COSMOS2020 and the CSP-I catalog utilized were the Kron aperture magnitudes named \textquote{MAG\_AUTO}; while the S82-MGC catalog utilized asinh magnitudes.

The cross-matching procedure of  SN~Ia hosts galaxies began by taking the Pantheon+ SN coordinates and using them to locate the corresponding sources in the COSMOS2020, S82-MGC, and CSP-I catalogs. Hosts were matched either by name (when available) or by selecting galaxies within 1 arcsecond of each Pantheon+ position.  For the datasets we downloaded and compiled ourselves, we searched within a 6 arcminute radius for SDSS-DR16 and a 0.4 arcminute radius for HST images from the MAST server, and selected the top three closest matches. We then visually inspected the images to check the SN~Ia and galaxy positions and confirmed that the correct host galaxy was matched in both the SDSS and HST catalogs. Their magnitudes were obtained from the SDSS catalog directly, whereas for HST we ran \texttt{Source Extractor} 
\citep{Bertin1996} on the images, adopted the \texttt{KRON\_RADIUS}-based apertures, manually inspected them to ensure consistency with the SDSS apertures, and used \texttt{MAG\_AUTO} for the final magnitudes.

In all, we gathered 36 crossmatches from COSMOS2020, 299 from the S82-MGM, and 97 galaxies for the CSP-I catalog. This resulted in a total of 431 galaxies, and after removing duplicate sources across the three catalogs, we obtained 292 unique galaxies. We add to this 13 hosts from \emph{HST}/WFC3-IR archival imaging, ensuring by inspection that the data is both non-saturated and consistent catalog optical/NIR host photometry. In total we analyze 312 galaxies SEDs with optical and NIR photometry, up from 97 from previous Pantheon+ work, and spanning redshifts $0 < z < 0.7$.

 \subsection{Catalog Zero Point Bias Correction}
 \label{subsec:cuts}
To ensure consistent photometry for SED fitting, and correct for systematic offsets in photometric calibration and aperture corrections, we tested the derived SEDs in each catalog for biases, or zero-point calibrations \citep{Ilbert2006, Ilbert2010,Dahlen2010ApJ...724..425D}. The zero-point calibration computes the gross residuals between observed photometry and the photometry expected by the best-fitting \texttt{prospector} models, and derives offsets for each passband of each catalog. For this analysis we extend our source sample to include galaxies which did not host SNe, to 4,436 sources from COSMOS2020, 94,407 from S82-MGC, and 84 from CSP-I catalogs that (a were flagged as ``good quality'' with no source extraction flags (b) had spectroscopic redshifts for precise zeropoint-corrections and the \texttt{EAZY} template fit did not have more than 3 data points with residuals over 5$\sigma$. This ensures to work only with galaxies that have reliable fits (see Table~\ref{tab:selection_cuts}). 

\begin{deluxetable}{l c c c}
\tablecaption{Selections for Zero-Point Corrections
\label{tab:selection_cuts}}
\tablehead{\colhead{Catalog}
& \colhead{Sources} & \colhead{ $z_{sp}$ + Good Phot.} & \colhead{Resid. Cut}
}
\startdata
COSMOS2020& 1.7 M & 6,620 & 4436 \\
S82-MGC & 15.3 M &  95,219 & 94,407  \\
CSP-I & 99&  -- & 84 \\
\enddata
\tablecomments{ The \textquote{$z_{sp}$ + Good Phot.} column indicates whether the data have an associated spectroscopic redshift and are flagged as high quality. The \textquote{Resd. Cut} column corresponds to the residual cut applied to the \texttt{EAZY} fits; further details are provided in Section~\ref{subsec:cuts}.}
\end{deluxetable}

Subsequently, we ran the photometric redshift code \texttt{EAZY} \citep{EAZY2008} on the photometry of these galaxies. The code fits template spectral energy distributions (SEDs) to the multi-band data and yields photometric redshifts. We ensured that these photo-z's were consistent with their respective spectroscopic redshift and measured the residuals from the data to the best-fit models. This zero-point test was not applied to the data we downloaded from SDSS-DR16 and MAST, as we only had a total number of 13 galaxies, which is too small to conduct a systematic zero-point correction.  We computed the mean residuals for each photometric band and applied these values as zero-point offsets to the corresponding filters.  These resulting offsets for each catalog are shown in Appendix~\ref{sec:zpbc}.

\section{Galaxy Parameterization and Stellar Mass Measurements}
\label{sec:stellarmass}
 
The effect of neglecting the near- and mid-IR wavelengths on stellar mass measurements, in the SEDs fits, has been studied in the literature \citep{Pforr2012, Pforr2013, Mobasher2015}. The addition of the IR data provide strong constraints on stellar masses from low-mass stars, which are less affected by dust extinction and re-emission. From \cite{Pforr2012}, it was shown that at z $\sim$ 0.5 the stellar mass can be underestimated by as much as $\sim$ 0.6 dex (at fixed IMF) with optical data alone and assuming exponentially declining SFH models. As mentioned above, it is unclear whether the presence of this \textquote{mass-step} is due to a physical relation or a limitation on tools and data; it is therefore crucial that we recalculate these masses from Pantheon+. We use the stellar population code \texttt{Prospector} \citep{Prospector2021} to robustly constrain stellar masses and stellar histories among other parameters~\citep{Leja2019, Leja2020}.

\begin{figure}
    \centering
    \hspace{-0.5cm}
    \includegraphics[width=1.0\linewidth]{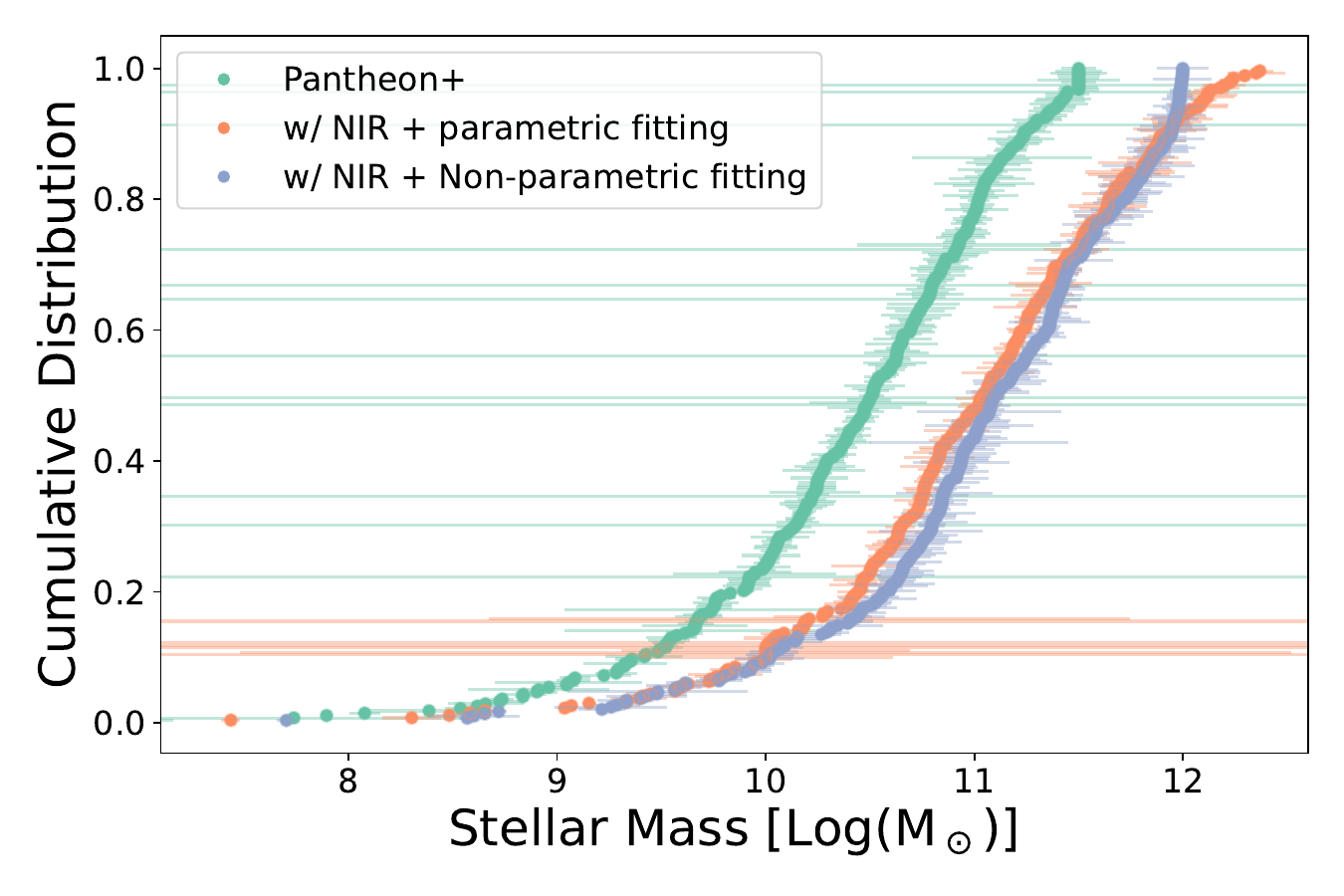}
    \caption{Cumulative distribution of derived masses, with mass errors, from the Pantheon+ sample (in green), and from the analysis in this paper (in blue). Also shown (in orange) is the effect of including NIR data and applying parametric fits with \texttt{prospector} illustrating the fraction of improvement from the NIR data alone. }
    \label{fig:pantheon_comp}
\end{figure}
          
Most SED fitting results assume a functional form for the SFH, with the most common choice being exponentially declining models (parametric). However, there are drawbacks to this, as they are shown to underestimate stellar masses for dusty and starburst galaxies \citep{Leja2019, Leja2020, Lower2020}. Therefore, we employ a non-parametric fit approach to our sample, which uses a piecewise constant to fit the SFR within specified age bins. This allows for accurate representation and flexibility in the model to capture complex and more realistic SFHs. 

In Figure \ref{fig:pantheon_comp} we illustrate the effect on derived stellar masses on a sample where parametric SFHs are applied to a largely optical data set (in green), where near-IR data are then included (in red), and finally where non-parametric fitting are applied to optical+near-IR dataset (in blue). Most in the shift to higher masses is actualized by including near-IR, and the small remainder is attributed to the additional flexibility of the non-parametric fitting. The weighted mean masses shift from $\log(M_{\odot})=10.72 \pm0.38$ (with a mean error of 0.63) for the Pantheon+ sample to $\log(M_{\odot})=11.17\pm0.66$ (with a mean error of 0.10) for the non-parametric with IR sample, respectively. Furthermore, we also compare the errors for the non-parametric model (blue) and the parametric SFH (red) as shown in Figure \ref{fig:pantheon_comp}. The median mass uncertainty for Pantheon+ is 0.0875 ($\sigma \sim$ 4.1) dex, whereas for our non-parametric is 0.081 ($\sigma \sim$ 0.0680) dex.  The errors have a much lower standard deviation in the non-parametric than the Pantheon+ sample, showing a stark improvement in mass uncertainties. 

\begin{figure*}
    \centering
    \hspace*{1cm}\includegraphics[width=0.51\linewidth]{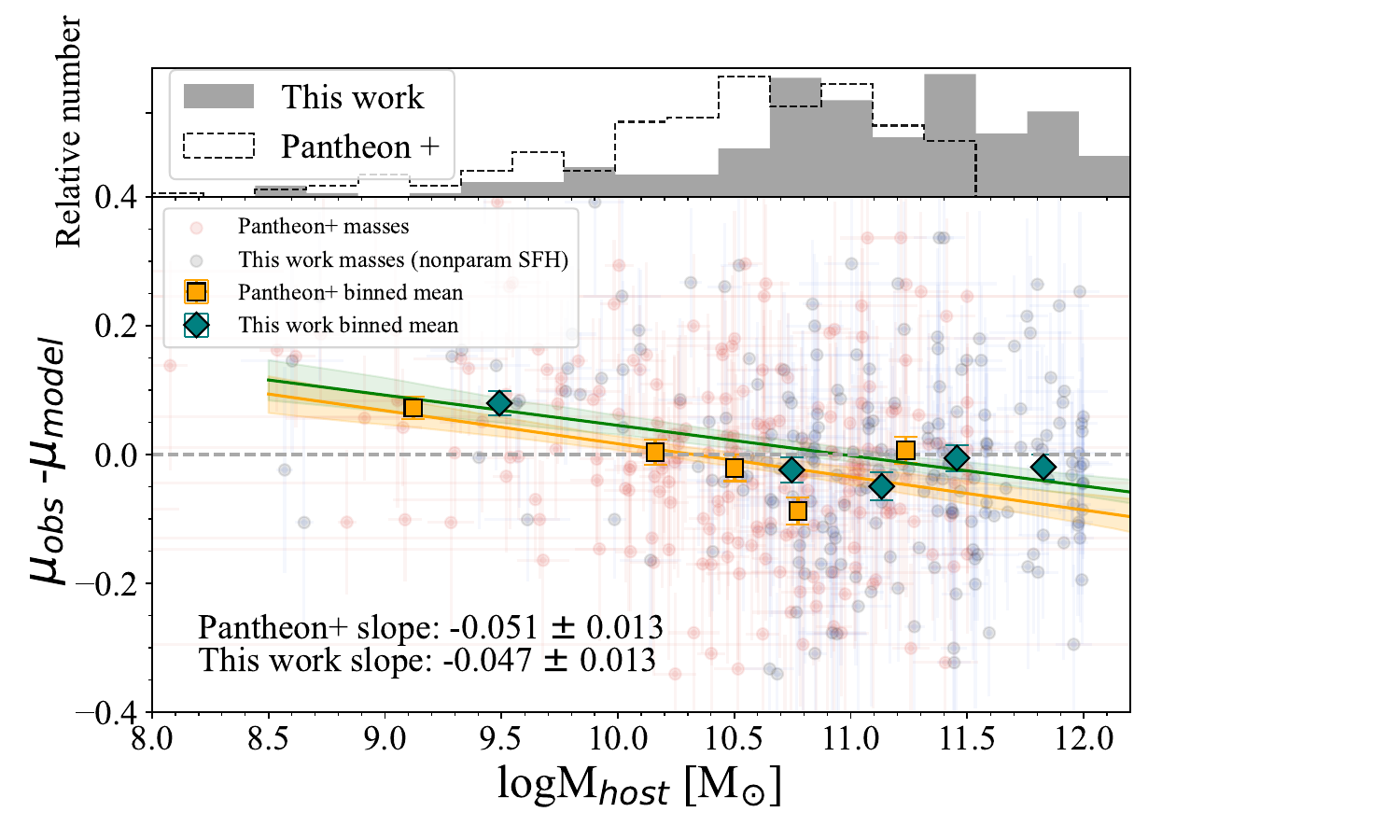}
    \hspace{-2.9cm}
    \hspace*{1cm}\includegraphics[width=0.51\linewidth]{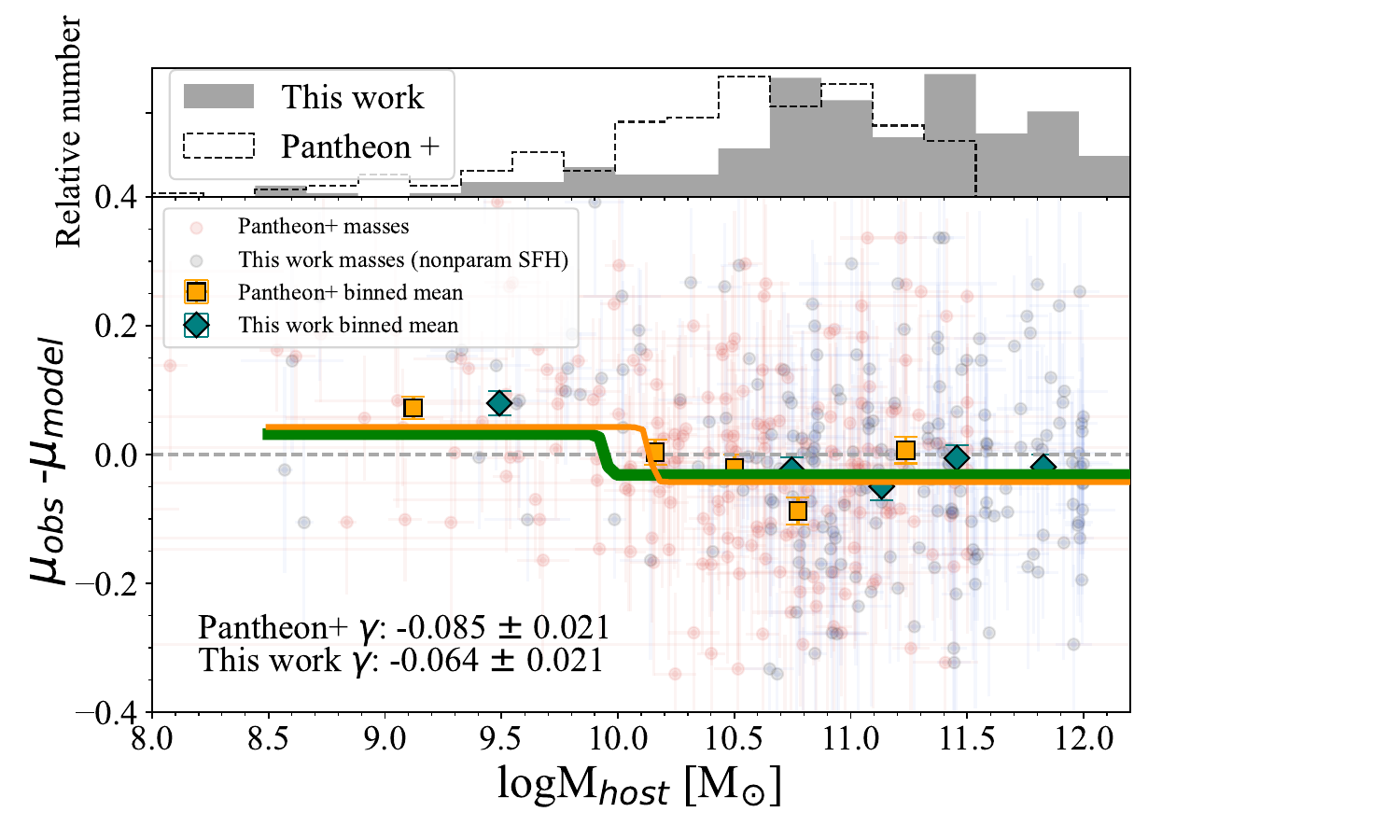}
    \caption{Left: Hubble residuals with the host stellar mass, where blue are the newly remeasured host masses, and red points the original measures from Pantheon+. Yellow and green squares are equilibrium-binned data for each, respectively. Yellow and green lines are the trends in each respective sample. The inset above shows the shift in the mass histograms. Right: Same as left, replacing linear fit with function defined by Equation~\ref{delta_mu}.}
    \label{fig:HR}
\end{figure*}

\section{TRENDS WITH THE HUBBLE RESIDUAL}
\label{sec:trends_hr}
\subsection{The Mass Step}
Figure~\ref{fig:HR} shows the SN HRs as a function of galaxy stellar mass for the 312 host galaxies remeasured in this study (light blue points), compared to the original measures in Pantheon+ (light red points). The square data points are the weighted mean $\log(M_{\odot})$ bins, for Pantheon+ (yellow) and this work (green), where each bin contains the same number of SNe. Following \cite{Kelly2010} and \cite{Sullivan2010}, we fit a line to the $\mu_{obs}-\mu_{model}$ and $\log(M_{\odot})$ data of Figure~\ref{fig:HR}, computing a slope of $-0.047\pm0.013$ for our non-parametric data (green line) which is slightly flatter than the slope for the original mass estimates ($-0.051\pm0.013$, yellow line in the left panel of figure). Following \cite{Scolnic2021}, one can fit the data with a sigmoid function of the form following Equation~\ref{delta_mu}, representing more of a step than a continuous trend. Both the linear slopes and step parameters are tabulated in Table~\ref{tab:gamma_values} in comparison to literature values. The shifts in masses as exemplified in Figure~\ref{fig:pantheon_comp} is also shown in the shift of these populations in Figure~\ref{fig:HR}, and in the inset histogram atop the figure. 

We explore the relationships of masses to other SN~Ia fitted parameters in Appendix~\ref{sec:hrr}.

\begin{deluxetable*}{l c c c }
\tablecaption{Literature mass-step relation 
\label{tab:gamma_values}}
\tablehead{\colhead{Work}
& \colhead{ Slope } & \colhead{ $|\gamma|$ } & \colhead{$\gamma_{err}$} 
}
\startdata
This Work & $-0.047 \pm 0.013$ & 0.064 & 0.021  \\
\cite{Scolnic2021} & $-0.051 \pm 0.013$ & 0.06 & --  \\
CSP-I & $\sim -0.05$ & $-0.074$ -- $-0.147$ & 0.030 -- 0.040 \\
\cite{Kelly2010} & $-0.15 \pm 0.05$ & 0.11 &  --\\
\cite{Sullivan2010} & $\sim -0.18$& $0.06$--$0.08$ & 0.028 -- 0.069 \\
\enddata
\end{deluxetable*}

\begin{figure}
    \centering
    \includegraphics[width=1.05\linewidth]{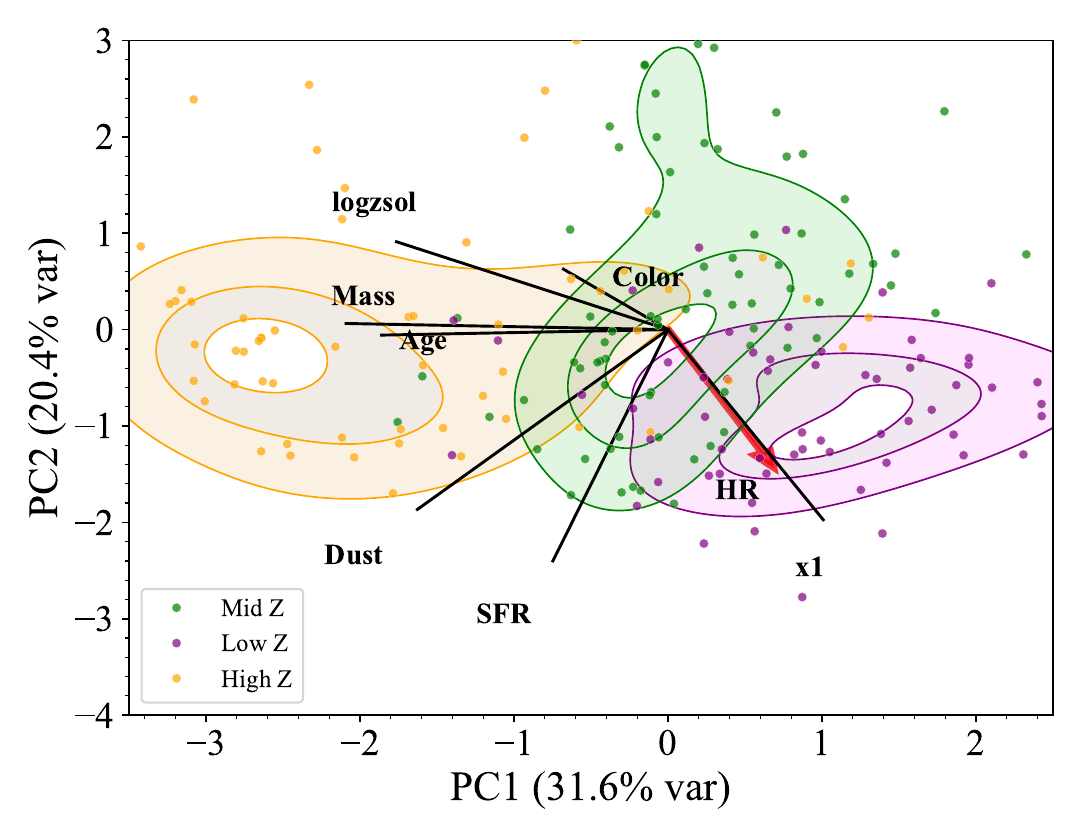}
        \caption{Principal component analysis of the host environmental parameters for our SNe~Ia sample. The HR shows the strongest negative correlation with stellar metallicity (logzsol) and the strongest positive correlation to stretch (x1).}
        \label{fig:pca}    
\end{figure}

\subsection{HR relations to other fitted galaxy parameters}
 Studies have found that brighter SNe~Ia are preferentially found in lower metallicity galaxies and within younger stellar populations \citep[e.g.,][]{Gallagher2008, Howell2009, Neill2009,Johansson2013}, pointing to a potential relation in the host mass$-$SN luminosity relation with respect to host metallicity. Some studies additionally argued that progenitor age is the dominant factor, with lower mass and higher sSFR galaxies producing the younger and brighter SN~Ia explosions~\citep{childress2013, Rigault2013}.  
 
 We explore these environmental relationships with the Hubble residuals via principal component analysis (PCA) to transform the high-dimensional data into a lower dimension to identify correlations. The first principal component (PC1) is the direction in space along which the data points have the highest or most variance, while PC2 accounts for the next highest variance in the dataset and must be uncorrelated with PC1. In Figure~\ref{fig:pca}, we show the Pantheon+ supernova characteristics and host parameters scattered with their respective probability density distributions. 

 The features used as inputs for the PCA analysis are SFR, dust, HR, mass, metallicity (logzsol), stretch (x1), color (c), and age. From the figure, we see lines of different lengths, representing the strength of the correlation of each feature. The direction determines whether they are positively or negatively correlated. From Figure \ref{fig:pca}, we see metallicity is the strongest negatively correlated with the HR feature, while x1 is the most positively correlated (as the line orientations are the closest in direction). Interestingly, although mass is negatively correlated with HR, it is not as strongly correlated as x1 or metallicity. Furthermore, the HR trend points toward low-metallicity galaxies, suggesting that higher residuals are associated with lower metallicities and higher stretch values. We also find that color is negatively correlated with HR; however, this correlation is relatively weak, as indicated by the much shorter vector length compared to features such as metallicity.


\subsection{HR variations with Metallicity and Stretch } \label{subsection:x1_metallicity}

\begin{figure*}
    \centering
        \includegraphics[width=0.75\linewidth]{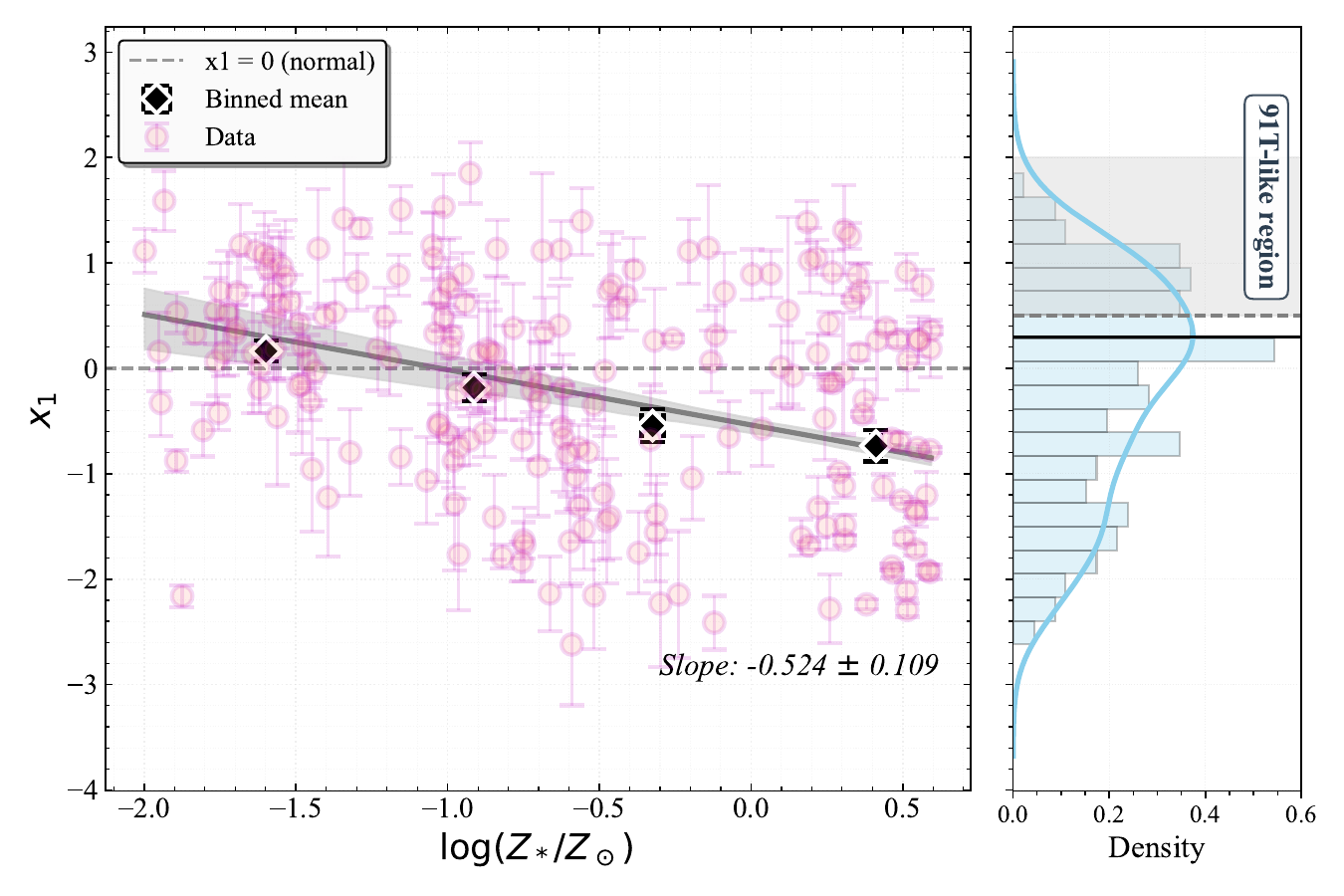}
        \caption{Stretch against host stellar metallicity. We notice a mildly strong negative correlation and that low metallicity tends to produce brighter SNe~Ia (higher x1). The right panel illustrates the stretch distribution and highlights how the 91T-like region largely overlaps with the stretch cuts we applied ($x1 > 0.3$), as discussed in Section~\ref{subsection:x1_metallicity}, where we showed that the mass-step trend effectively disappears once these high-stretch events are removed.}
        \label{fig:x1_Z}
\end{figure*}

Based on the results of the PCA, we further test the effects of host metallicity and SN stretch on the Hubble residuals. In Figure \ref{fig:x1_Z}, we plot these two parameters and show that stretch and the stellar metallicity are indeed negative correlated, as indicated by the PCA, suggesting that low metallicity galaxies generally produce brighter SNe~Ia in this analysis. We also tested the mass-step for metallicity and stretch under different cuts to explore how the trend behaves.

\begin{figure*}
        \centering
        \includegraphics[width=0.9\linewidth]{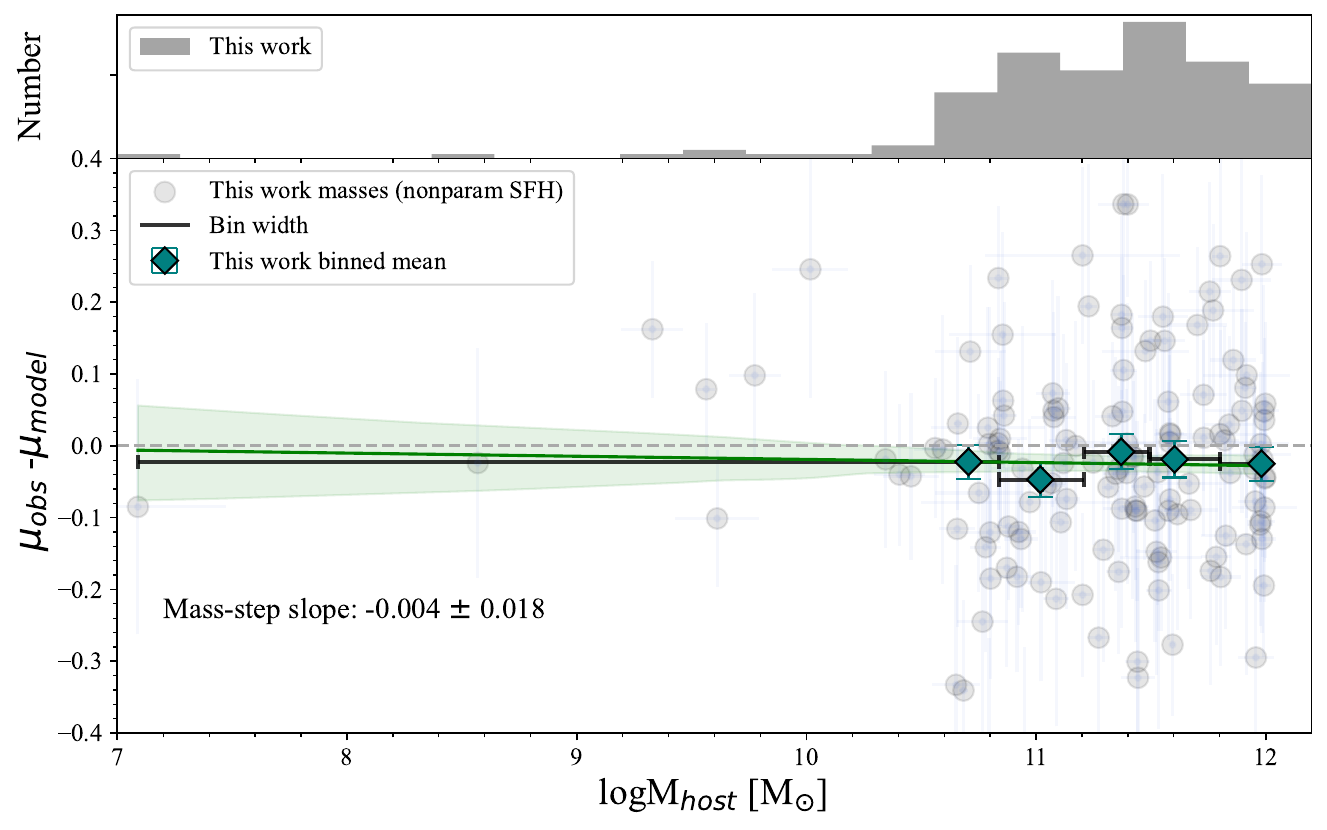}
        \caption{Hubble residuals against host mass excluding low-metallicity hosts. We find the trend to effectively disappear after the removal. }
        \label{fig:HR_lowz}
\end{figure*}

We tested the \textquote{mass-step} relationship with different metallicities and split the sample into 3 subgroups: low metallicity (Z $<$ $-1$), intermediate metallicity ($-1$ $<$ Z $<$ 0), and high metallicity (Z $>$1). We chose -1 the low metallicity value is well below the mean metallicity in the local ($z<0.3$) universe~\citep[$Z\approx0.1$, ][]{Gallazi2005}. We removed the lowest metallicity subsample ($Z<$ $-1$), and redid the mass-luminosity analysis, as shown in Figure \ref{fig:HR_lowz}. We note that the mass-step effect is greatly diminished, with a slope ($-0.004\pm0.018$) highly consistent with zero. This could point to low metallically causing much of the trend and the need for a separate treatment for these galaxies. 

Similarly, we found that cutting $x1 > 0.3$ also made a significant difference in the slope of the HR relation with mass (see Appendix~\ref{sec:hrr}). We also note that low mass galaxies of this sample almost exclusively host high stretch ($x1 > 0.3$) SNe~Ia. Furthermore, it is interesting that this region in stretch overlap with typical stretches of 99aa and 91T-like SNe~Ia~\citep{Burgaz2025}, as is also shown in Figure \ref{fig:x1_Z} right, and are frequently attributed to low mass star-forming galaxies~\citep{Hammuy1996,Hamuy2000,Sullivan2006,Neill2009, Phillips2024}. 

Overall, we found a strong relationship between low metallicity, stretch, and the mass-step. Although it not surprising that high stretch could be correlated with low-metallicity galaxies~\citep{Timmes2003}, it is surprising to see how effective the mass-step disappears after removing these two parameters independently. In both cases, we see a very similar behavior, when removing these atypical environments or properties. However, it is worth noting that after removing these low metallicity and high stretch SNe~Ia and host data, not much of the data remains below masses of $10^{10} M_{\odot}$. This scarcity of low-mass hosts could itself be responsible for the observed trend. Specially given that low mass galaxies appear to strictly produce brighter SNe~Ia (Figure \ref{fig:x1_Z}).

\section{Effects on Cosmological Constraints}
\label{sec:cosmology}
\begin{figure}
    \centering
    \includegraphics[width=0.99\linewidth]{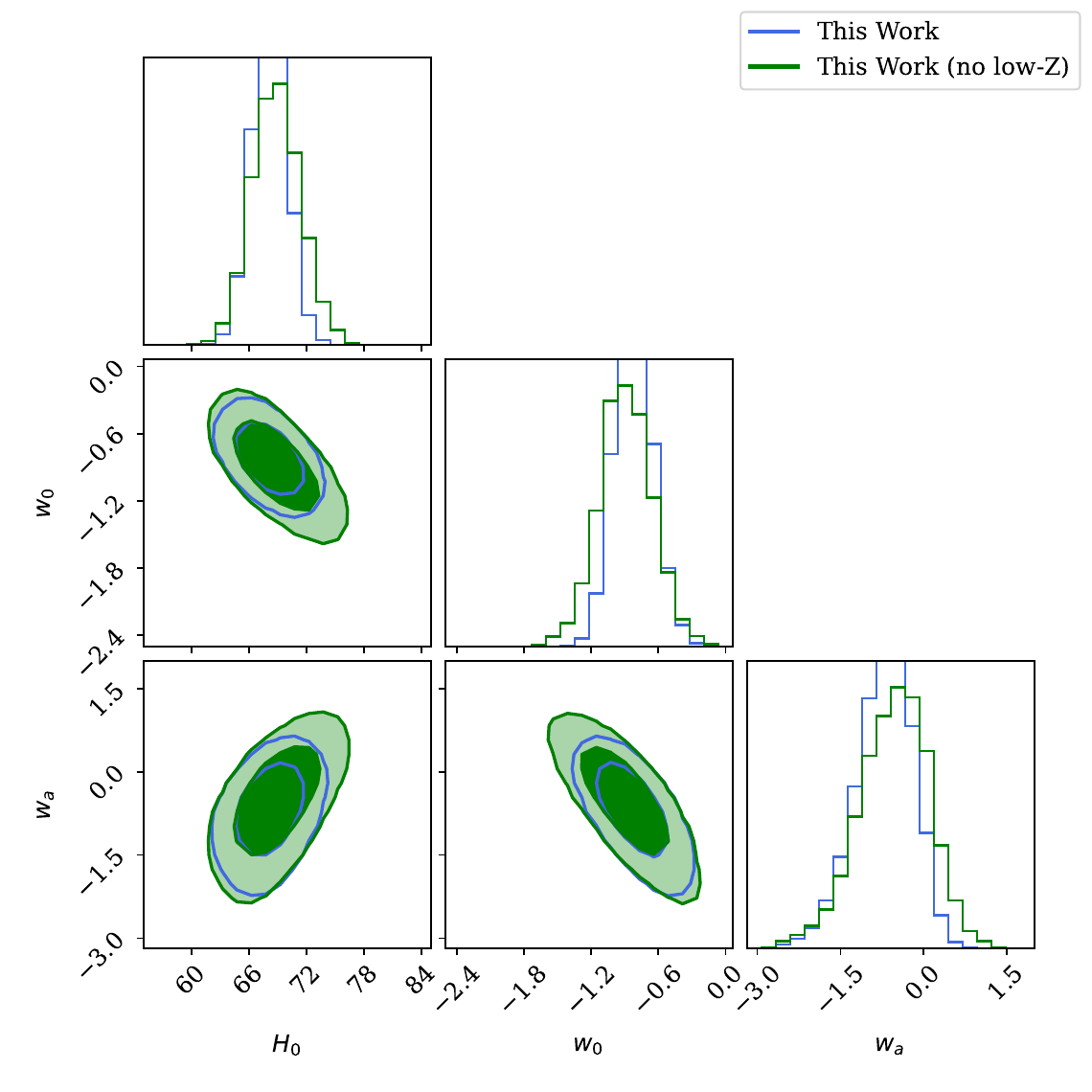}
    \caption{Corner plot of the probability distribution of the hubble residuals, $w_{0}$, and $w_{a}$. The blue distribution is obtained from the SNe~Ia sample and constraints from the bao and plack data, whereas the blue is the same, but the the SNe~Ia sample excludes SNe~Ia whose host galaxies are low metallicity.}
    \label{fig:cosmology_fits}
\end{figure}

 We found that low-metallicity galaxies and stretch as a result, were the dominant contributors for the observed \textquote{mass-step}. While this step has been empirically corrected for in previous analysis, we tested whether excluding the galaxies with low metallicity  (Z $<$ -1.0 Z$_{\odot}$) could produce a measurable shift in the inferred Hubble constant $H_{0}$. To quantify this, we ran cosmological parameter inference using the \texttt{cobaya} sampler \citep{Torrado2019} in two configurations: (1) the galaxies in our sample and (2) excluding galaxies with low metallicity (effectively correcting for the mass-step). We sampled three primary cosmological parameters:  the Hubble constant $H_{0}$ in the range between 60 to 80 km/s/Mpc, the baryon density $w_{a}$ between -3 to 2 , and the dark energy equation of state today $w_{0}$ between $-2.5$ to 0. The spatial curvature $\Omega_{k}$  known as the curvature density parameter was fixed to 0. The power spectrum was created from one of the theory codes \texttt{CLASS} supported by \texttt{cobaya}. Additionally, we added the  \cite{Planck2015} temperature, polarization, and lensing likelihoods with BAO2024 \citep{BAO2024}, and our SNe~Ia sample to jointly constrain the relations. In Figure \ref{fig:cosmology_fits}, we see the corner plots of the two distributions, the sample (blue) and the corrected \textquote{mass-step} sample in green. In the full sample, we see $w_{0}$ = -0.81 $\pm$ 0.2, which is consistent with $\Lambda_{CDM}$ ($w_{0}$ = $-1$) at $\sim$1$\sigma$. The  $w_{a}$ = -0.71 $\pm$ 0.5 suggests mild evolution, but also consistent with 0 at $\sim$1.3$\sigma$. However, when we exclude the low metallicity data, we notice a shift toward more negative $w_{0}$ (toward the cosmological constant) and less negative  $w_{a}$ ( toward no evolution), both less than 1$\sigma$ away from $\Lambda_{CDM}$ (0.5$\sigma$ and 0.9$\sigma$, respectively). The values we found are listed in table \ref{cosmo_results}. This suggests that the low-metallicity subsample drives the constraints away from $\Lambda_{CDM}$: their inclusion shifts $w_{0}$ by $+0.07$ (0.4$\sigma$, further from -1) and $w_{a}$ by $-0.17$ (0.3$\sigma$, further from 0).

\begin{deluxetable}{l c c c}
\tablecaption{Cosmology fitted parameters
\label{tab:cosmology_results}}
\tablehead{\colhead{Data} & \colhead{$H_0$}
& \colhead{$w_{0}$} & \colhead{$w_{a}$} 
}
\startdata
Pantheon+, full sample& $67.41^{+0.52}_{-0.84}$ & $-0.84 \pm 0.06$ & $-0.65^{+0.28}_{-0.32}$\\
~~~w/ mass-step & & &\\
This subsample (312),& 68.11 $\pm$ 1.8 & $-0.81$$\pm$ 0.2 & $-0.71$ $\pm$ 0.5\\
~~~w/ no mass-step& & &\\
no Low Z &68.85  $\pm$ 2.1 & $-0.89$ $\pm$  0.2 & $-0.55$ $\pm$ 0.6 \\
\enddata
\tablecomments{All three fits were applied to our dataset, with varying data cuts as described in Section~\ref{sec:cosmology}, in order to assess their impact on the cosmological results.}
\label{cosmo_results}
\end{deluxetable}

\begin{figure}
    \hspace{-0.8cm}
    \centering    \includegraphics[width=1.08\linewidth]{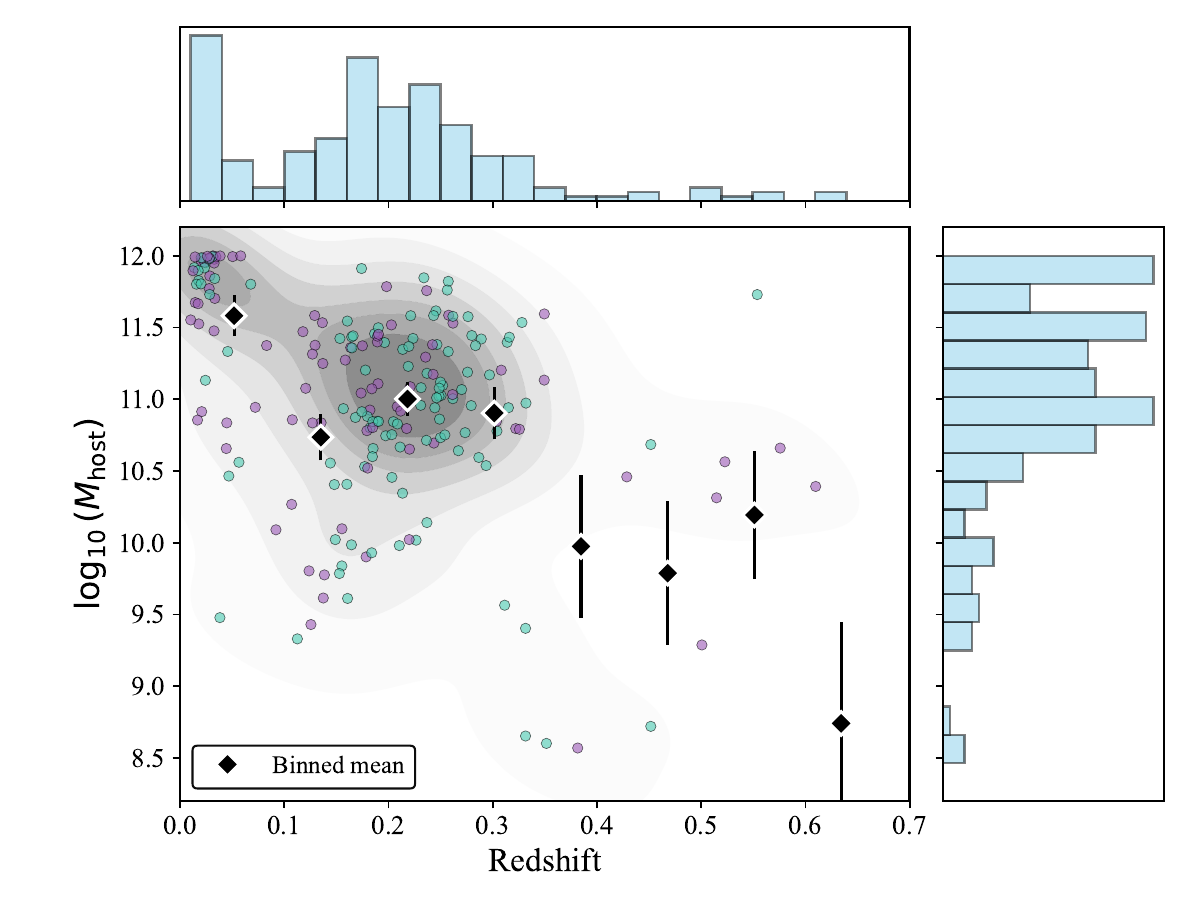}
    \caption{Mass vs redshift, showing that the higher the redshift the smaller the mass. The purple points are SNe~Ia brighter then -19.2 mag and the cyan are dimmer than -19.2 mag. This plot shows that SNe~Ia brightness do not depend on the redshift. }
    \label{fig:m_z_relation}
\end{figure}

\section{Discussion and Conclusions}
\label{sec:conclusion}
We gathered available optical and NIR data for galaxies overlapping in the Pantheon+ sample. We improve on host galaxy stellar mass measures by using a sample with complete $ubvrizJH$
coverage, and utilizing non-parametric SFH fitting techniques. From our analyses, we saw that previous studies had indeed underestimated the masses by a factor of atleast $\sim$ 2 in linear scale. Although the mass-step relation was largely unaffected, we did see a large shift in the mass estimates, with important implications for host-environmental studies. 


We further investigated the relation between SN~Ia luminosity, host-galaxy stellar mass, and redshift as shown in Figure \ref{fig:m_z_relation}. We do not find an obvious trend in the brightness of SNe~Ia as a function of the redshift: both brighter SNe (the purple color data points) and dimmer (in cyan) span the entire redshift range. However, we do observe a strong decline in host galaxy mass at higher redshifts, qualitative consistent with galaxy evolution expectations. However, the trend is much steeper than what is expected from literature results. This suggests a potential selection biases which could be driven by selection quality cuts, but unclear.

In exploring the Hubble residual (taken from Pantheon+) correlation against various physical parameters, including: SFR, dust, mass, age, and metallicity, we found metallicity to be the most directly negatively correlated. Inferring that the HR increases with smaller metallicities host environments. We separated the sample into three different subclasses, low (Z $< Z_{\odot}$), median (0 $<$ Z $<$ $Z_{\odot}$) and high (Z $>$ $Z_{\odot}$). We removed the low-metallicty sample and found the \textquote{mass-step} trend to significantly diminish from -0.047 mag/log$M_{\odot}$ to -0.004 mag/log$M_{\odot}$. Similarly, we found a similar behavior with stretch. When removing the data with high stretch ($x1 > 0.3$), the \textquote{mass-step} also significantly flatted from -0.047 mag/log$M_{\odot}$ to -0.009 mag/log$M_{\odot}$. This points to metallicity being a dominant factor in the mass-step relationship in our study, as well as indicating a potential connection between low-metallicity and high-stretch. We note that all of the estimated parameters in this work are derived from global host properties, rather than the local properties, which could vary the galaxy properties. Nevertheless, previous studies have found that local and global host properties yield broadly consistent trends for SN~Ia standardization \citep[e.g.,][]{Jones2018}. Therefore, while local measurements may introduce additional scatter, we do not expect them to qualitatively alter the primary trends or conclusions presented here.

While the physical origin of these correlations remains uncertain, there is an observed relationship between SN~Ia properties and host-galaxy environment. When focusing on the most common subclasses of Type Ia supernovae normal (SNe~Ia)—namely, 91T-like, and 91bg-like events—it is observed that SNe in lower-mass host galaxies (which are typically of later morphological types or have higher specific star formation rates) are, on average, brighter and slower to evolve \citep[higher stretch, x1 e.g.][]{Hammuy1996,Howell2001,Gallagher2008, Neill2009, Gonzalez-Gaitan_2011, Gonzalez-Gaitan_2014, Li2011a}. 91T-like SNe are found in star-forming galaxies, and none have been discovered in elliptical galaxies, where the stellar population is predominantly old \citep[e.g.][]{Howell2001,Gallagher2008, Li2011a}. 

In contrast, 91bg-like events are more common in elliptical and lenticular (E-S0) galaxies, though they occasionally occur in early-type spirals \citep[e.g.][]{Howell2001, Li2011a}. Normal SNe~Ia are found in host galaxies of all morphological types, from ellipticals to late-type spirals (e.g., Li et al. 2011a). The range of metallicities in main-sequence stars that evolve into white dwarfs could theoretically influence the amount of  $\rm{^{56}Ni}$ produced in SNe~Ia, thereby affecting their properties, such as luminosity and decline rates. \cite{Timmes2003} suggested that less luminous SNe~Ia originate from high-metallicity progenitors that produce less $\rm{^{56}Ni}$. 


While it is challenging to ascertain whether stellar population age or metallicity is the primary factor behind these correlations, several studies suggest that progenitor age is the key determinant. Lower mass galaxies with higher specific star formation rates (sSFR) tend to produce younger SNe~Ia, implying shorter delay times are associated with more massive white dwarfs and more energetic explosions, leading to higher luminosity \citep{childress2013,Johansson2013, Rigault2013}. 

On the other hand, theoretical models suggest that metallicity also influences the outcome: lower-metallicity progenitors produce larger amounts of radioactive  $\rm{^{56}Ni}$ and thus brighter SNe~Ia \citep{Timmes2003}. Finally, it is important to recognize that age and metallicity are inherently coupled through galactic and chemical evolution. Younger stellar populations tend to reside in low-mass, metal-poor, and actively star-forming galaxies, whereas older populations are generally found in massive, metal-rich systems.

This work reveals a significant correlation between the mass-step and host-galaxy metallicity. When low-metallicity galaxies are removed from the sample, the slope of the mass-step relation changes markedly, suggesting that metallicity plays an important role in driving this effect.Additionally, this removal shifts the inferred cosmological parameters closer to $\Lambda_{CDM}$, suggesting that low-metallicity environments may have different intrinsic properties that bias cosmological measurements if not properly accounted for. However, we caution that the current sample is imbalanced, with relatively fewer low-mass, metal-poor galaxies, which may bias the inferred trend. Future work should aim to obtain deeper near-infrared observations for these low-mass systems to improve completeness and test whether the observed metallicity dependence of the mass-step persists.

\begin{acknowledgements}
The following software packages were used in the analysis presented in this work: numpy \citep{numpy}, matplotlib \citep{matplotlib}, scipy \citep{scipy}, astropy \citep{astropy}. This research has made use of NASA’s Astrophysics Data System.
\end{acknowledgements}

\clearpage
\appendix

\section{Catalog Zero Point Bias Corrections}\label{sec:zpbc}
\begin{figure}[H]
    \centering
    \includegraphics[width=0.65\linewidth]{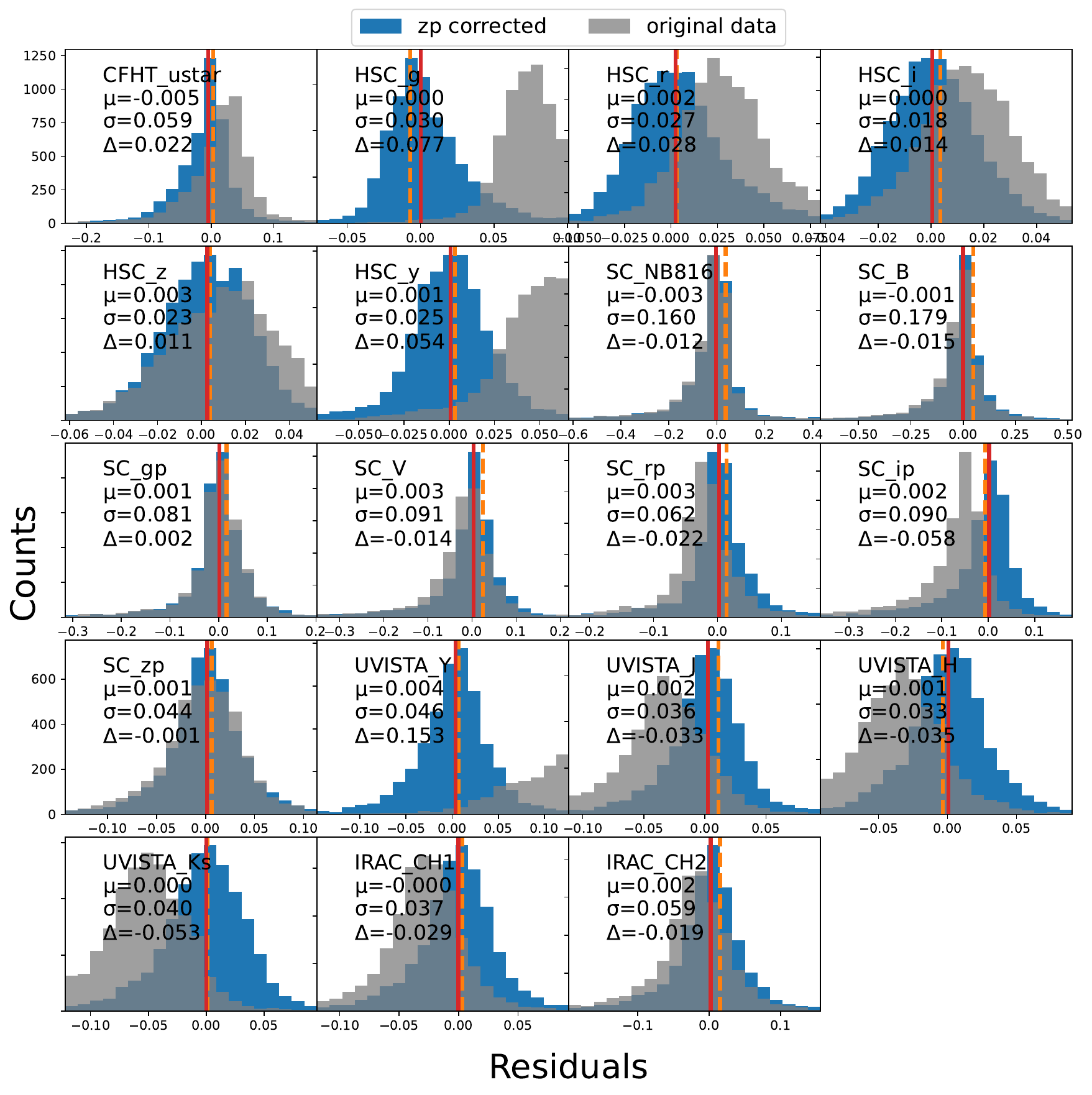}
    \caption{Histogram of the COSMOS2020 high quality data displaying the residuals from the host galaxy SED fits from the data. The histogram in gray shows the original residuals before the correction and in blue after the correction. The red line is the mean, while the orange dotted line represents the mode. The $\mu$ and $\sigma$ represents the mean and standard deviation of the corrected histogram. The $\Delta$ represents the systematic offset needed to correct for (i.e the zero-point calibration) }
    \label{fig:cosmos_zp}
\end{figure}

\begin{figure}[H]
    \centering
    \includegraphics[width=0.45\linewidth]{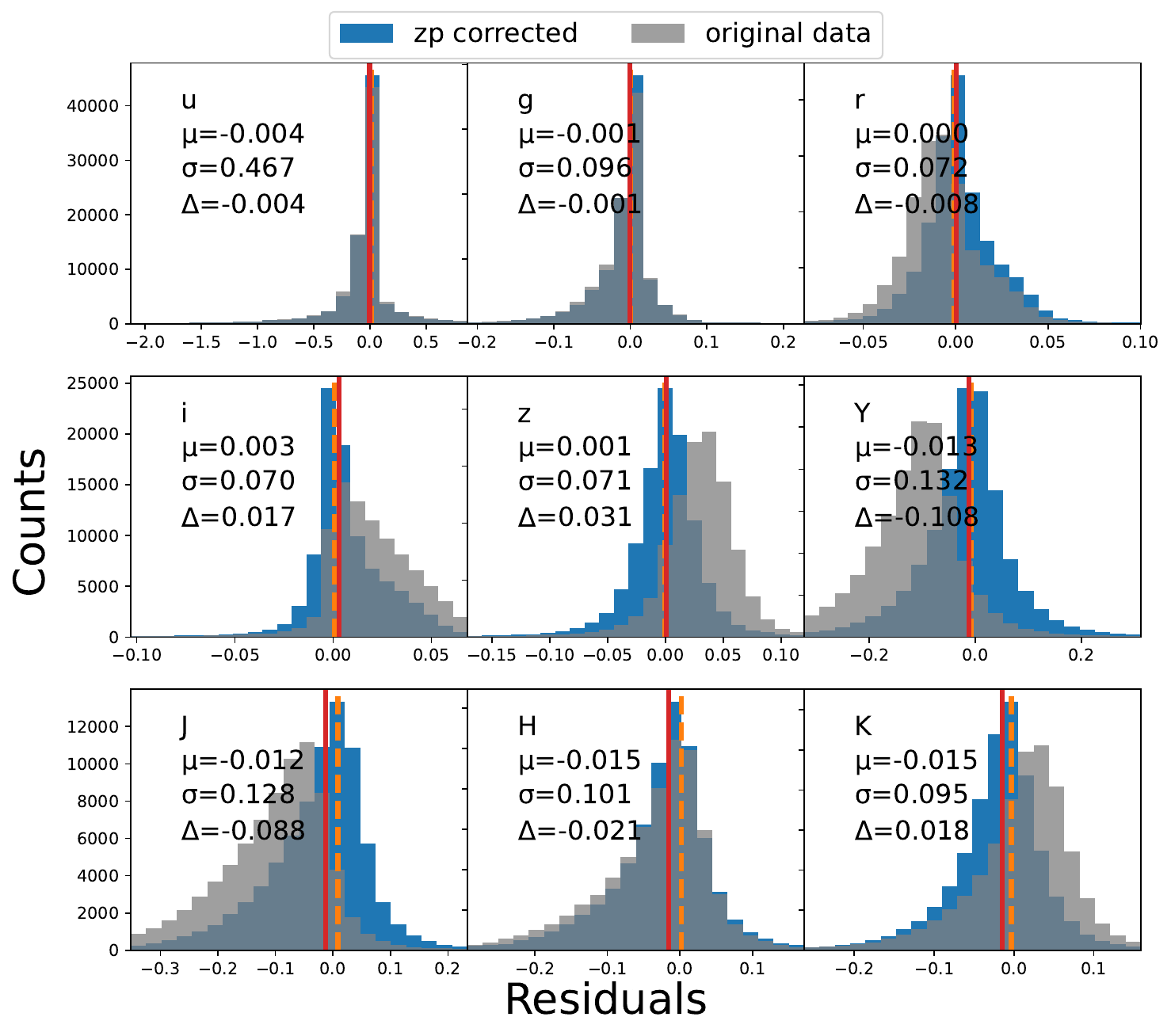}
    \includegraphics[width=0.45\linewidth]{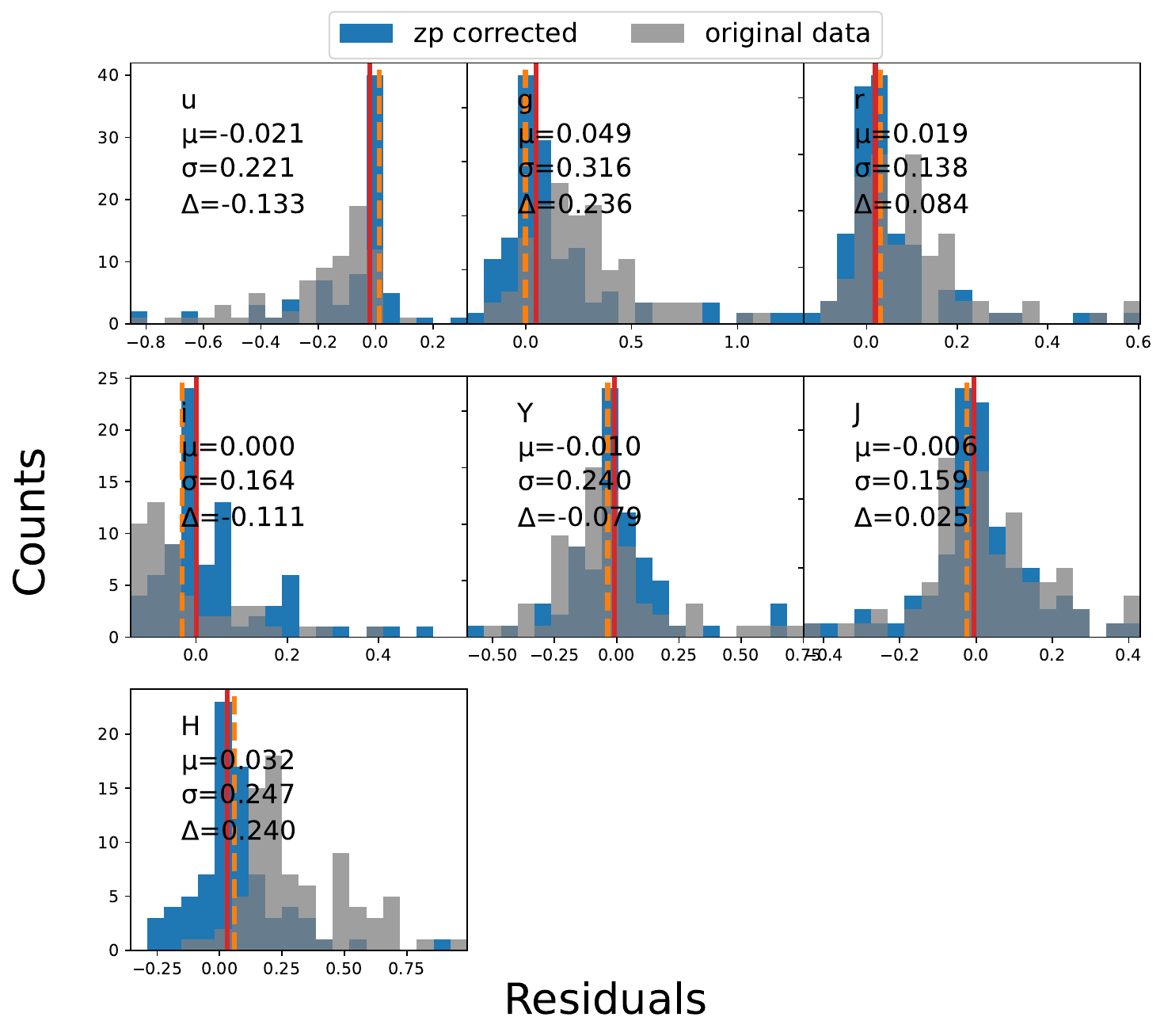}
    \caption{Same as Figure~\ref{fig:cosmos_zp} but for the S82-MGC (\textit{left}) and CSP-I (\textit{right}) high quality data.}
    \label{fig:sdss_zp}
\end{figure}

Figures~\ref{fig:cosmos_zp} and \ref{fig:sdss_zp} show the effect of zero point bias corrections for each catalog used in this study. Each figure presents two histograms per filter: the gray histograms show the mean residuals between \texttt{prospector} fits and catalog photometry before applying the correction, and the blue histograms show the residuals afterward. The corrected distributions are centered near zero, indicating that the applied offsets effectively remove the systematic differences between the data and the model photometry.

\section{Hubble Residuals Relationships}\label{sec:hrr}

\begin{figure}[H]
    \centering
     \includegraphics[width=0.65\linewidth]{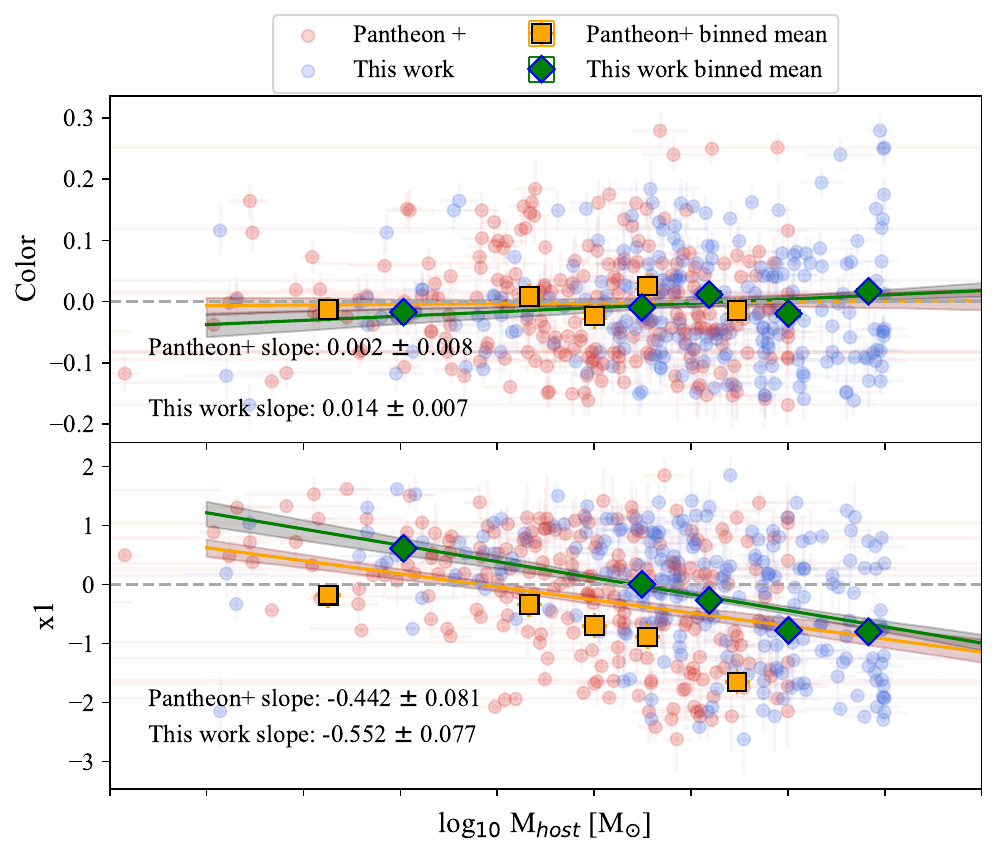}
    \caption{Color and stretch ($x_1$) relationship vs mass, stretch appears to be significantly correlated with mass, whereas color does not appear largely affected.}
    \label{fig:x1_M}
\end{figure}

We explored relationships between the SNe~Ia fitted parameters from Pantheon+ and the galaxy fitted parameters inferred from \texttt{prospector}. In Figure~\ref{fig:x1_M} we show the $x_1$ and color parameters against host galaxy mass. We note there is a significant trend in $x_1$ with mass for the Pantheon+ sample (-0.44) and our sample (-0.55). The color parameter on the other hand, had no significant correlation on either the Pantheon+ (0.002) or our sample (0.014).  It is also worth noting that low mass galaxies appear to strictly produce SNe~Ia with positive stretch, which are often associated to brighter SNe~Ia \citep{Johansson2013,Gallagher2008, Howell2009, Neill2009}. However, here the $x_1$-mass trend likely arises as a correlative secondary effect driven by the relationships between $x_1$ and metallicity ($Z$) as shown in Section~\ref{subsection:x1_metallicity}, and the known galaxy mass-metallicity relation~\citep[e.g., ][]{Tremonti2004}.

\begin{figure}[H]
\centering
    \includegraphics[width=0.48\linewidth]{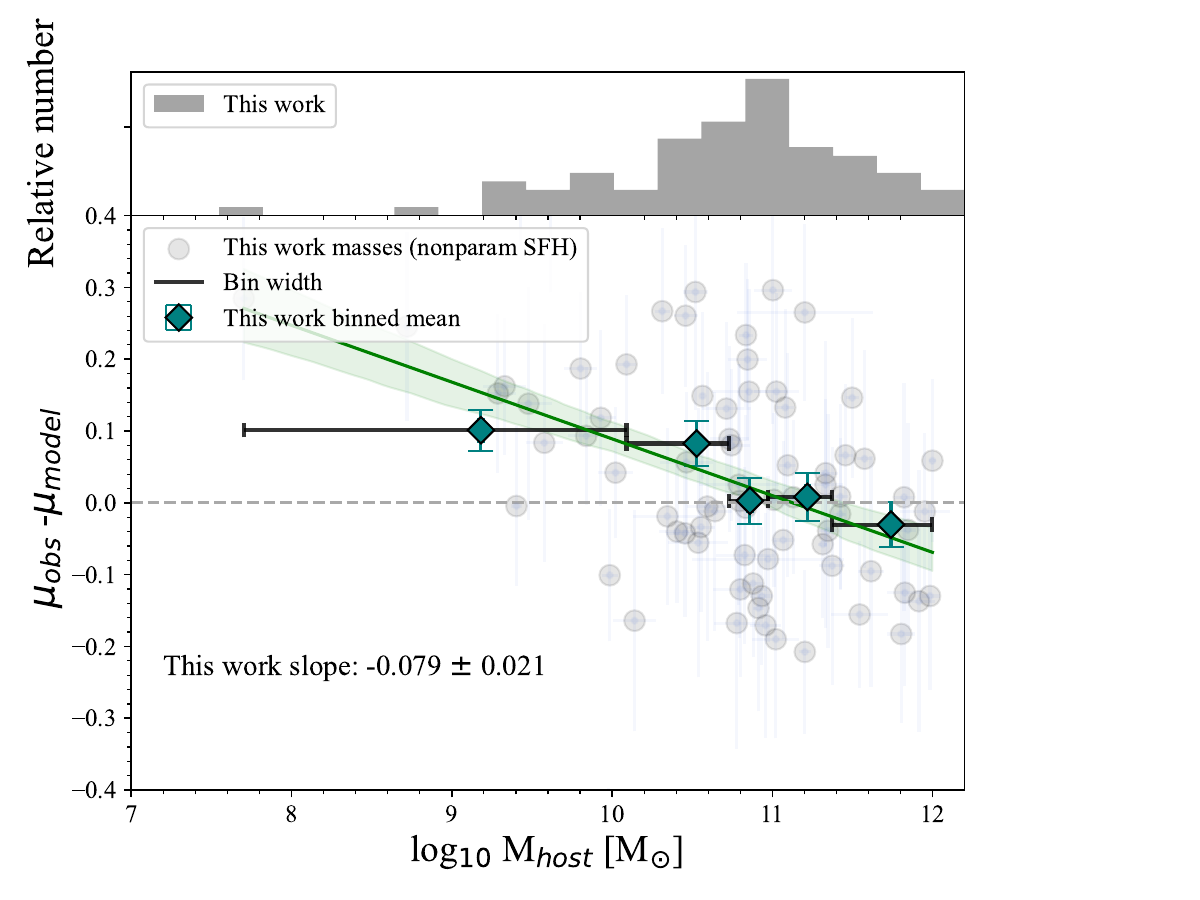}
    \includegraphics[width=0.48\linewidth]{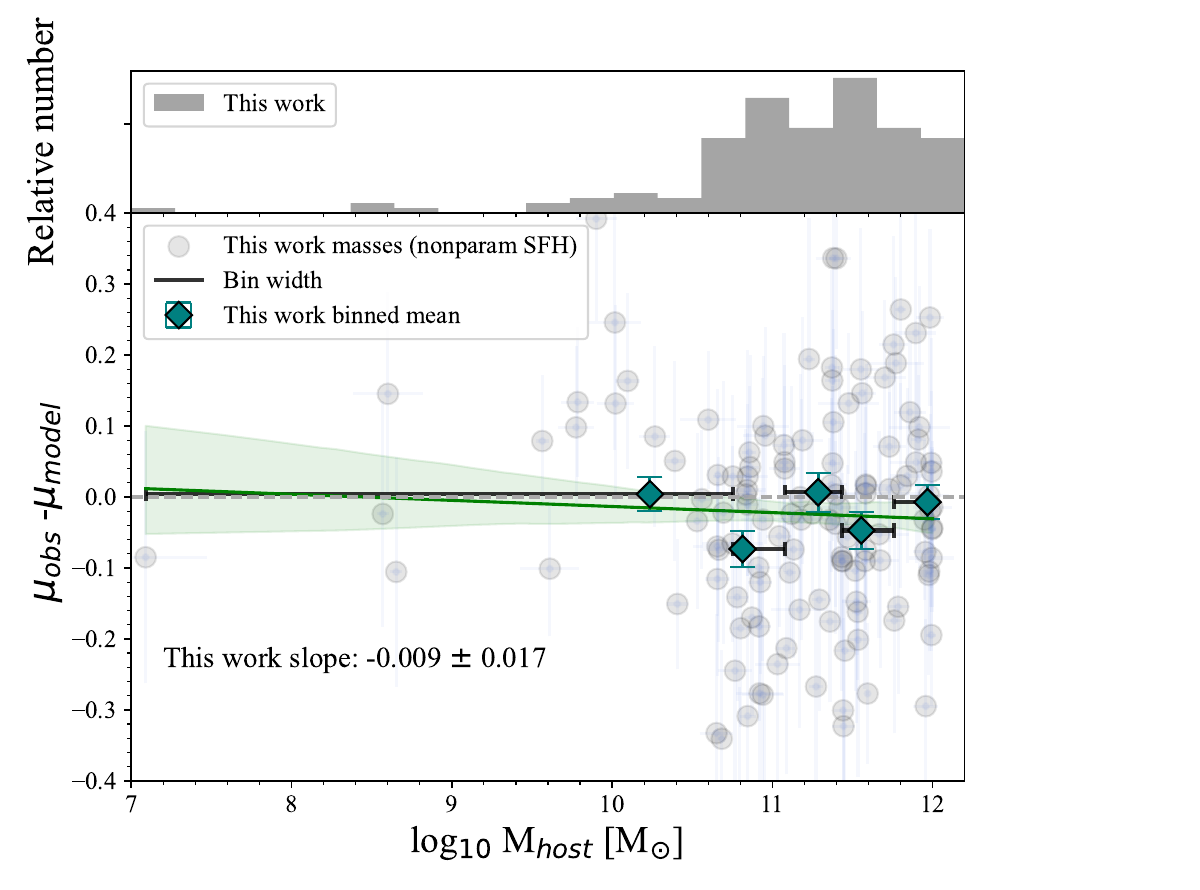}
\caption{Left: Hubble residuals against mass for high stretch SNe~Ia ($x1 > 0.3$) and their host galaxies. Right: same as left but for $x_1\le0.3$.}
\label{fig:HR_x1}
\end{figure}

This is further illustrated in Figure~\ref{fig:HR_x1}, which shows the trend of the Hubble residuals against mass when using high stretch as a selection criteria. The left panel shows the trend when only selecting SNe~Ia with $x_1 > 0.3$, whereas the right excludes high-stretch SNe, focusing on $x_1\le0.3$. We notice a strong correlation in the mass-step in the high-$x_1$ sample. However, once that data was removed, the slope became consistent with zero.

\clearpage
\bibliographystyle{aasjournal}
\bibliography{references}

\end{document}